\def\spose#1{\hbox to 0pt{#1\hss}}

\def\multleft#1{\hbox to size{\vbox {\halign {\lft{##}\cr #1}}\hfill}\par}
\def\multright#1{\hbox to size{\vbox {\halign {\rt{##}\cr #1}}\hfill}\par}

\def\today{\ifcase\month\or January\or February\or March\or April\or May\or
      June\or July\or August\or September\or October\or November\or December\fi
      \space\number\day, \number\year}
\def\s{\hbox{\phantom{5}}}	%one space
\def\cm{{\rm\thinspace cm}}

\def\erg{{\rm\thinspace erg}}

\def\K{{\rm\thinspace K}}

\def\km{{\rm\thinspace km}}
\def\kpc{{\rm\thinspace kpc}}

\def\Mpc{{\rm\thinspace Mpc}}
\def\Msun{\hbox{$\rm\thinspace M_{\odot}$}}

\def\s{{\rm\thinspace s}}

\def\yr{{\rm\thinspace yr}}

\def\ergpcmsqps{\hbox{$\erg\cm^{-2}\s^{-1}\,$}}

\def\ergps{\hbox{$\erg\s^{-1}\,$}}

\def\kmps{\hbox{$\km\s^{-1}\,$}}

\def\pcm{\hbox{$\cm^{-3}\,$}}

\def\psqcm{\hbox{$\cm^{-2}\,$}}

\def\kmpspMpc{\hbox{$\kmps\Mpc^{-1}$}}

\def\H2{\hbox{H$_{2}$}}
\def\Lya{Ly$\alpha$}

\documentclass[usegraphicx]{mn2e}

\usepackage{times}
\usepackage{amssymb}
\include{defn}

\begin{document}
\hsize=6truein

\title[HI in the proto-cluster environment at $z>2$: absorbing haloes and the \Lya~forest]{HI in the proto-cluster environment at $z>2$: absorbing haloes and the \Lya~forest \thanks{Based on observations performed at the European Southern Observatory, Chile (Programme ID: 71.B-0036(A))}}
\author[R.J.~Wilman, M.J.~Jarvis, H.J.A.~R\"{o}ttgering and L. Binette]
{\parbox[]{6.in} {R.J.~Wilman$^{1}$\thanks{Email: r.j.wilman@durham.ac.uk}, M.J.~Jarvis$^{2}$, H.J.A.~R\"{o}ttgering$^{3}$ and L.~Binette$^{4}$\\ \\
\footnotesize
1. Department of Physics, South Road, University of Durham, Durham, DH1 3LE. \\ 
2. Astrophysics, Department of Physics, Keble Road, Oxford, OX1 3LE. \\
3. Sterrewacht Leiden, Postbus 9513, 2300 RA Leiden, the Netherlands. \\
4. Instituto de Astronomia, UNAM, Ap.70-264, 04510 Mexico, DF, Mexico \\}}
\maketitle

\begin{abstract}
We present VLT-UVES echelle spectra of the \Lya~emission and absorption in five radio 
galaxies at redshifts z=2.55--4.1. Together with data from our pilot study, we 
have a sample of 7 such systems with radio source sizes $\sim 1-90$\kpc~with which to address 
the origin of the absorbing gas. Echelle resolution again reveals that some systems 
with $N_{\rm{HI}}>10^{18}$\psqcm~in lower resolution data in fact consist of several weaker absorbers with 
$N_{\rm{HI}}<10^{15}$\psqcm. We identify two groups of HI absorbers: strong absorbers with $N_{\rm{HI}} 
\simeq 10^{18}-10^{20}$\psqcm~and weaker systems with $N_{\rm{HI}} \simeq 10^{13}-10^{15}$\psqcm. There are 
none at intermediate $N_{\rm{HI}}$.

The strong absorbers may be a by-product of massive galaxy formation or could instead represent material cooling 
behind the expanding bow-shock of the radio jet, as simulated by Krause. New observations are required to discriminate 
between these possibilities. We argue that the weaker absorbers with $N_{\rm{HI}} \simeq 10^{13}-10^{15}$\psqcm~are 
part of the \Lya~forest, as their rate of incidence is within a factor of 2--4 of that in the IGM at large. Such column 
densities are consistent with models of a multi-phase proto-intracluster medium at $z>2$. 
\end{abstract}


\section{INTRODUCTION}
High-redshift ($z>2$) radio galaxies (HzRGs) are among our best probes of massive galaxy formation. 
The K-z relation demonstrates that their supermassive black holes ($10^{9}-10^{10}$\Msun) 
reside in the most massive elliptical galaxies at their epoch (e.g. Jarvis et al.~2001a; Willott et al.~2003). 
Measurements of galaxy clustering and of galaxy over-densities around individual sources, imply that they 
inhabit proto-cluster environments which will evolve into rich clusters at $z=0$ (for a review of radio 
source clustering and proto-clusters, see R\"{o}ttgering et al.~2003). 

Deep within these gravitational potential wells and surrounding the HzRGs to radii of tens to several hundred 
\kpc, are luminous ($>10^{44}$\ergps~in \Lya) emission line nebulae. Their properties bear witness to the 
complex gaseous processes in nascent massive galaxies, viz. gravitational collapse and shock heating, cooling, 
bursts of star formation and feedback in the form of superwinds; the latter can inject substantial quantities 
of energy and metals into the gas, enriching it to solar or super-solar metallicity out to tens of \kpc~(e.g. 
Villar-Martin et al.~2001). The presence of a powerful radio source has a profound impact on such an 
environment: within the dense interstellar medium of the galaxy, the relativistic radio jet drives shocks into 
the clumpy gas, leading to extreme kinematics ($FWHM > 1000$\kmps; Villar-Martin, Binette and Fosbury~1999), 
additional star formation and the emission line signatures of shock ionisation (Best et al.~2000; De Breuck et 
al.~2000). At radii beyond the radio source, the gas is kinematically quiescent (reflecting purely gravitational 
motions) and the continuum from the quasar nucleus is the main source of ionisation (Villar-Martin et al.~2003). 

New insight into the gaseous haloes of HzRGs came with the discovery by R\"{o}ttgering et al.~(1995) and 
van Ojik et al.~(1997) that the majority of small radio sources (those with projected linear size below 50\kpc) 
exhibit spatially resolved absorption in their \Lya~profiles, with HI column densities of $10^{18}-10^{19.5}$\psqcm. For 
the $z=2.92$ HzRG 0943--242, Binette et al.~(2000) discovered associated CIV$\lambda\lambda$1548,1551~absorption at the 
same redshift as the main \Lya~system. Through photoionisation modelling they argued that the absorbing gas is of lower 
metallicity ($Z \sim 0.01Z_{\rm{\odot}}$) and is situated beyond the emission line gas, outside the high pressure radio 
source cocoon. This material, they claimed, is a relic reservoir of low metallicity, low density ($n \sim 10^{-2.5}$\pcm) 
gas. In larger radio sources, the radio source cocoon plausibly engulfs and pressurises the outer halo entirely, causing 
it to be seen in emission rather than absorption, and thereby accounting for the anti-correlation between radio source 
size and the presence of HI absorption found by van Ojik et al.~(see also Jarvis et al.~2001b for the HI absorption:radio 
source size correlation).

With the Ultraviolet-Visible Echelle Spectrograph (UVES) at the VLT, we recently examined the absorbers in two HzRGs 
with over ten times the spectral resolution of R\"{o}ttgering et al. and van Ojik et al. Our targets were the 
aforementioned 0943--242, and 0200+015 ($z=2.23$) and the results were presented by Jarvis et al.~(2003) 
and  Wilman et al.~(2003). In 0943--242, the absorbers exhibit no additional structure -- the main absorber is 
still consistent with an HI column density of $10^{19}$\psqcm, in line with the Binette et al.~(2000) model. In contrast, 
a very different view of 0200+015 emerges: the single absorber with HI column density $\simeq 10^{19}$\psqcm seen at low 
resolution now splits into two $\sim 4 \times 10^{14}$\psqcm~systems; these extend by more than 15\kpc~to 
obscure addtional \Lya~emission coincident with a radio lobe and highly fragmented absorption is also seen on 
the red wing of the emission line at this position. The detection of one of the sub-systems in CIV suggests 
that the absorbing gas has undergone substantial metal enrichment, to perhaps as high as $10Z_{\rm{\odot}}$. 
Based on the smaller radio source size in 0943--242 (26\kpc~versus 43\kpc~for 0200+015), we conjectured that the radio 
source age (as inferred from its linear size) is the parameter controlling the evolution of: (i) the structure/kinematics 
of the absorbing halo, through interaction and shredding of the initially quiescent shells; (ii) its metallicity, 
through enrichment by starburst superwind triggered concurrently with the nuclear radio source. 

If the specific effects of the radio source can be understood, high resolution spectroscopy of such \Lya~haloes provides an opportunity to probe the \Lya~forest around high-redshift galaxies and hence to test theories for how such systems 
form out of the intergalactic medium (IGM). For quasars, such attempts are completely hampered by their strong UV continua, 
which through photoionisation locally reduce the number of \Lya~forest clouds below the IGM 
average (the so-called proximity effect). Lacking strong line of sight continua, radio galaxies offer a more direct
probe of the distribution of HI in such dense environments. Due to the nature of hierachical clustering, the opacity of 
the \Lya~forest is expected to increase in the vicinity of massive objects, and observations of background quasars do indeed
show excess HI absorption at radii of 1--5$h^{-1}$\Mpc~from $z \sim 3$ Lyman-break galaxies (LBGs) (Adelberger et al.~2003). 
However, within $0.5h^{-1}$\Mpc~there appears to be a deficit of HI, which they attribute to the influence of `superwind' 
outflows from the LBGs. From a theoretical stand-point it is not clear that superwinds alone can account for this observation, 
and additional effects such as local photoionization (Bruscoli et al.~2003) and the `filling in' of absorption with \Lya~emission 
from the LBG itself (Kollmeier et al.~2003) may be needed.

In order to address the origin and evolution of the absorbing shells and their connection with the \Lya~forest, we need 
to expand our sample of HzRGs with high resolution spectroscopy. For this reason, we have recently acquired UVES spectra for 
a further 5 HzRGs, yielding a sample of 7 HzRGs with radio source sizes in the range $\sim 1-90 $\kpc~(projected). Here we 
present the analysis and interpretation of the new dataset. The paper is structured as follows: in section 2, we describe the target selection, observations and data reduction, and in section 3 present the \Lya~profiles and absorption model fits on an 
object-by-object basis. In section 4 we compile numerous statistics for the absorption line ensemble and compare them with the 
properties of HI absorption in the IGM. In section 5, we interpret these results in the light of simulations of: (i) expanding 
radio sources in dense environments; (ii) the \Lya~forest around high-redshift galaxies. Section 6 contains our conclusions.

Throughout the paper we assume a cosmology with $H_{\rm{0}}=70$\kmpspMpc, $\Omega_{\rm{M}}=0.3$ and 
$\Omega_{\rm{\Lambda}}=0.7$.

\section{TARGET SELECTION, OBSERVATION AND DATA REDUCTION}
The targets were drawn from the list of $z>2$ radio galaxies compiled from the literature by De Breuck et al.~(2000). 
Once the galaxies invisible from the VLT at the observation epoch were excluded, we imposed the following selection criteria: 
firstly, that the galaxies be bright enough in \Lya~emission for echelle spectroscopy with integration times $\sim 3$ hours (slit 
fluxes of at least $7 \times 10^{-16}$\ergpcmsqps); secondly, that the sample comprise radio sources covering projected linear 
sizes $\sim 1$\kpc~to $\sim 100$\kpc~(5 of our 7 targets are in the 25--50\kpc~range, the range identified by Jarvis et al.~2003 as `critical' for metal enrichment). For all but one of the targets 
(1755--6916) moderate resolution ($\sim$ a few \AA) spectra exist from which the presence of some \Lya~HI absorption is inferred.
In the latter respect they are representative of the general radio galaxy population at these redshifts: we know from van Ojik 
et al.~(1997) that 90 per cent of the smaller ($<50$\kpc) exhibit such absorption, dropping to 25 per cent for larger sources; the 
analysis of De Breuck et al.~(2000) using the \Lya~asymmetry parameter offers qualitative support to this conclusion in a larger sample of lower resolution spectra (although the asymmetry parameter is insensitive to absorption at the peak of the underlying 
profile). Our targets have monochromatic radio source powers in the range log $P_{\rm{325 MHz}}=36.0 \pm 0.3$ 
(erg~s$^{-1}$~Hz$^{-1}$), uncorrelated with redshift. De Breuck et al.~(2000) report the existence of a weak correlation between
log $P_{\rm{325 MHz}}$ and \Lya~line luminosity, and our galaxies occupy a well-populated part of this plane. In summary, our targets 
are entirely representative of the known population of radio galaxies at these redshifts.

The new observations were performed with the UVES echelle spectrograph on VLT UT2 at the European Southern Observatory on the nights of 2003 July 26 and 27. For our targets we used only the red arm of the spectrograph, with cross-disperser 3 and a central wavelength of either 5200\AA, 5800\AA~or 6250\AA. On-chip binning of $2 \times 3$ (spatial $\times$ spectral) was used, resulting in a pixel size of 0.36~arcsec. The resulting pixel scales in the dispersion direction were in the range 0.05--0.06\AA, but spectra were often re-binned to a minimum of 0.1\AA~per pixel to improve the signal-to-noise ratio. The slit width was fixed at 1.2~arcsec and its position angle aligned with the radio axis. The observation log is shown in Table~1.

Offline data reduction was performed in IRAF, following the procedures for echelle data reduction described by Churchill~(1995). For each target, the sub-frames were bias-subtracted, median combined to remove cosmic rays and flat-fielded. The order-definition frames were used to determine the locations of the orders, which were then extracted and wavelength calibrated using a ThAr arc. The wavelength dependence of the blaze function of the grating was determined from the flat-field frame, and then removed from the data before the orders were combined into a single piece of spectrum in the region around redshifted \Lya. Unfortunately, we have no useful spectra of the CIV$\lambda\lambda1548,1550$ doublet for any of our objects, as the lines are either outside the wavelength range or too weak. Hence all our analysis concerns the \Lya~line.

\begin{table}
\caption{Log of observations}
\begin{tabular}{|llllll|} \hline
Target  & z &  Radio source              & Exposure & Slit PA & Seeing  \\
        &   &   size$^{\dagger}$ (\kpc)  & time (hr) & (deg) & (arcsec) \\ \hline
0121+1320 & 3.517 & 1.4 & 2.8 & 89 & 0.70 \\ 
1338-1942 & 4.106 & 39 & 4 & 161 & 0.55 \\
1545-234 & 2.755 & 48 & 2.5 & 159 & 0.70 \\
1755-6916 & 2.55 & 88 & 4.75 & 84 & 0.55 \\
2202+128 & 2.704 & 26 & 4 & 86 & 0.70 \\ \hline
\end{tabular}
$\dagger$~Projected linear size
\end{table}

\section{LYMAN ALPHA PROFILE FITTING} 
The \Lya~profiles were assumed to consist of a series of discrete absorption systems, modelled as Voigt profiles with free parameters of redshift, $z_{\rm{abs}}$, HI column density, $N_{\rm{HI}}$, and Doppler parameter, $b$,  atop an emission envelope. For the latter we use a gaussian, except in one galaxy where a lorenzian is clearly a much better fit. In fitting the profiles in this way, we assume that each absorber is in effect a screen between us and the emission source which simply scatters \Lya~photons away from our line of sight. This independence between emitter and absorber does, however, represent an idealised case since \Lya~is a resonant line. As such it can undergo multiple scatterings during its escape from an optically-thick medium, greatly increasing its path length and in the process rendering it extremely sensitive to any background opacity in the form of dust. This, and the frequency shifts which occur upon scattering, can in principle lead to very complex \Lya~profiles.

\begin{figure*}
\begin{centering}
\includegraphics[width=8cm,angle=0]{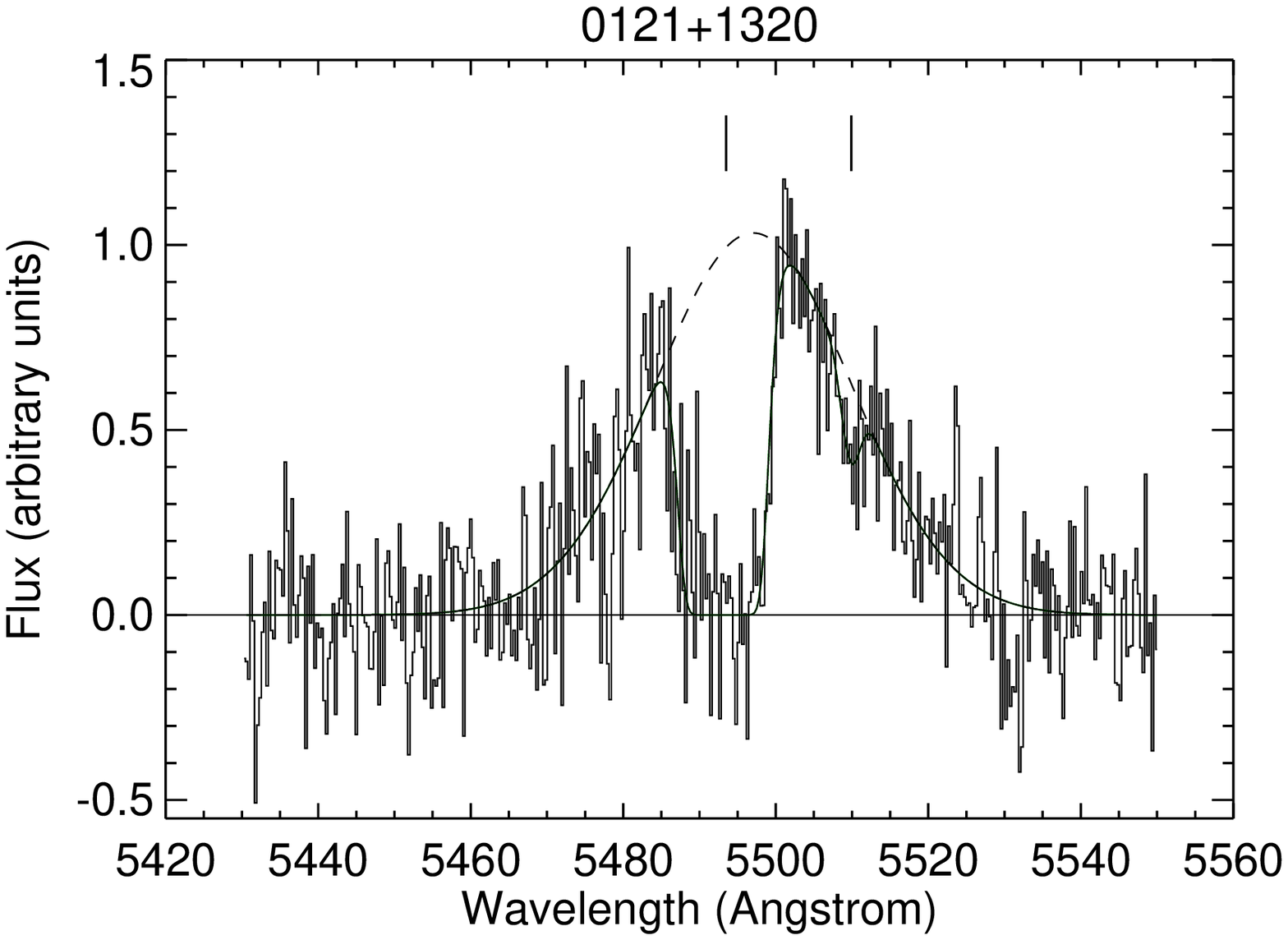}
\includegraphics[width=8cm,angle=0]{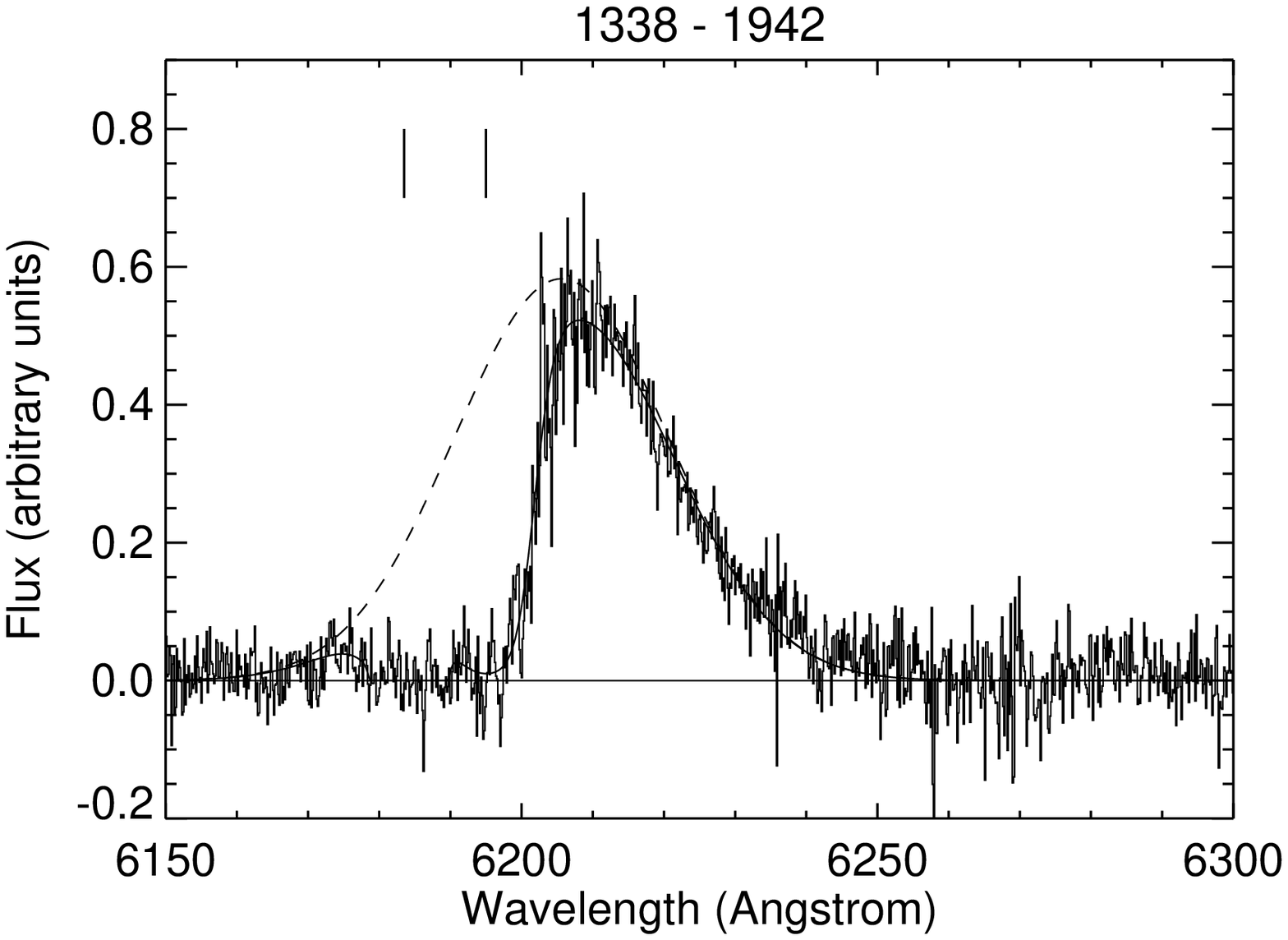}
\includegraphics[width=8cm,angle=0]{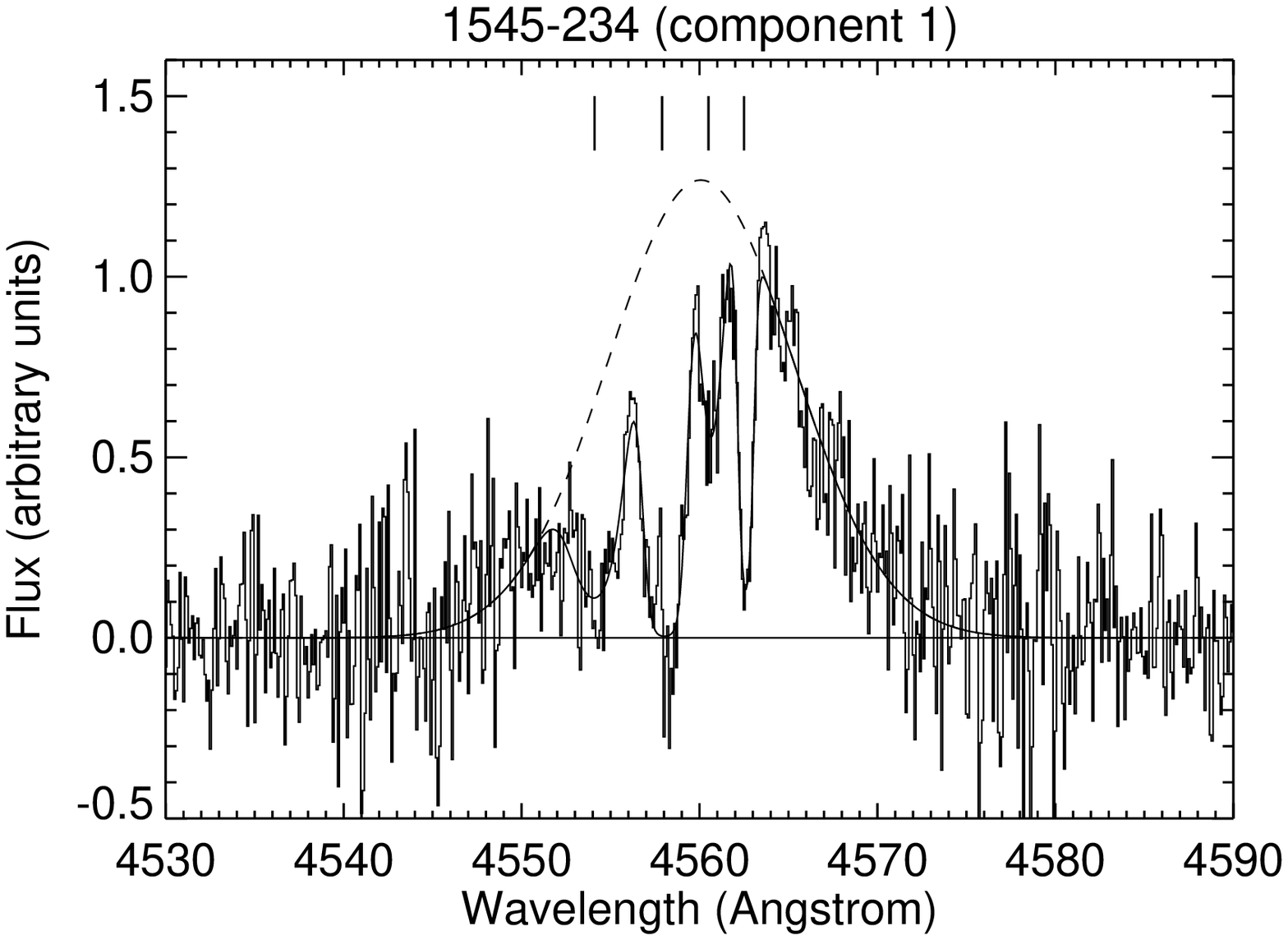}
\includegraphics[width=8cm,angle=0]{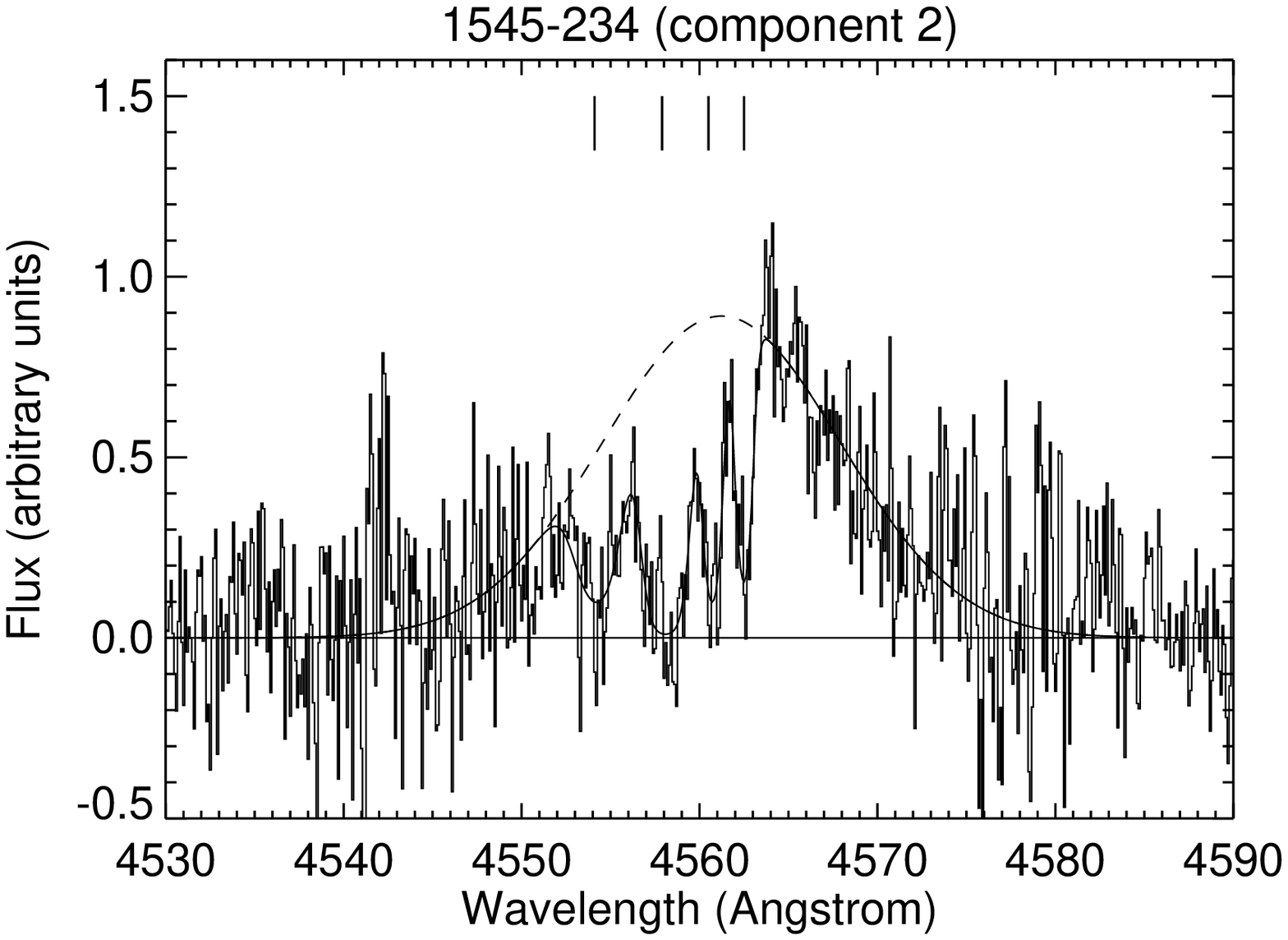}
\includegraphics[width=8cm,angle=0]{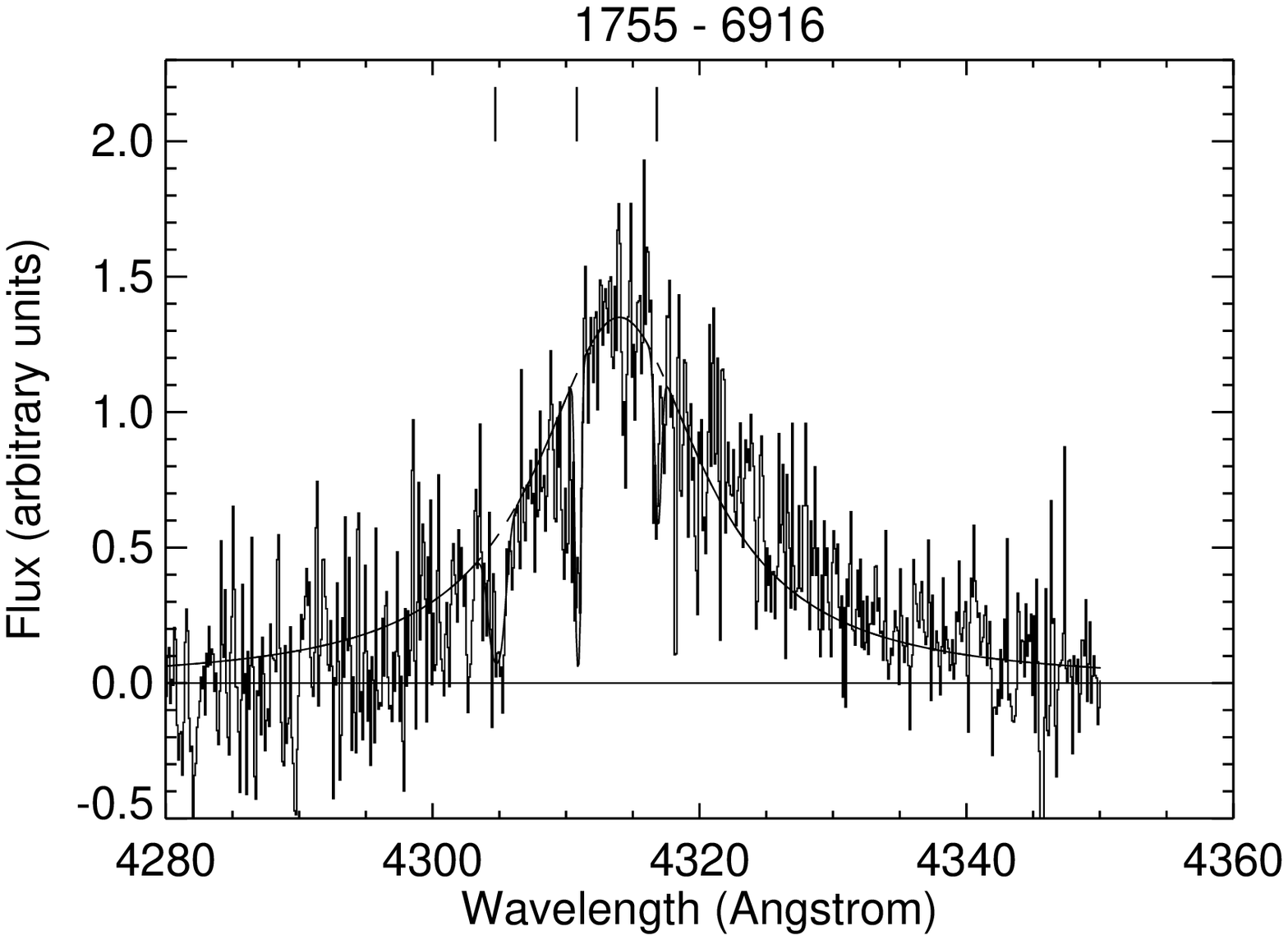}
\includegraphics[width=8cm,angle=0]{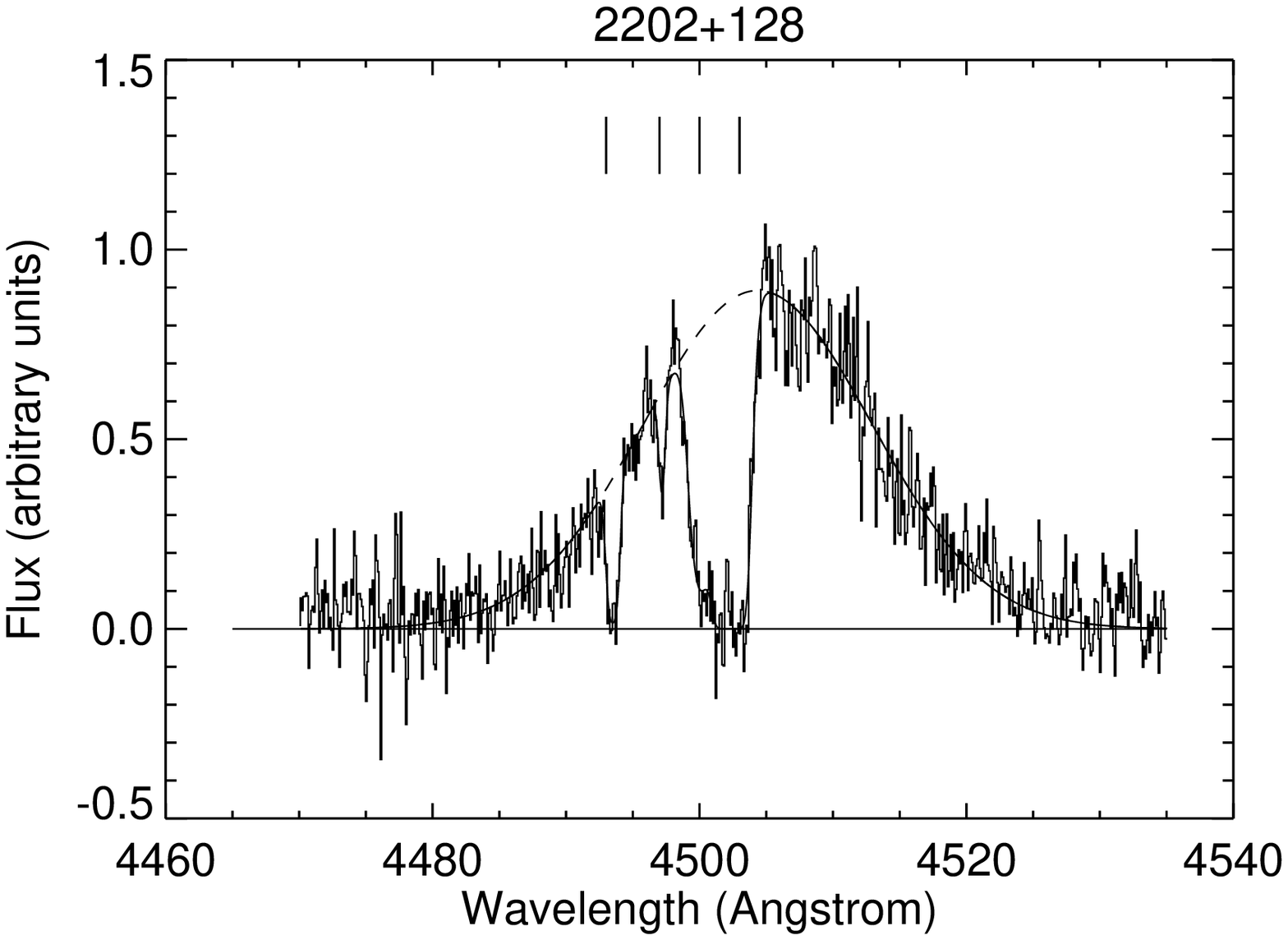}
\caption{Absorption model fits to the \Lya~emission line in each of our new radio galaxy targets. The dashed lines show the best
fit gaussian or (for the case of 1755--6916) lorenzian emission envelope, upon which a series of discrete absorption systems, 
modelled with Voigt profiles and indicated by vertical ticks, is superimposed. The fitted parameters are given in Table~2. For 
1545--234 we show spectra for each of the two spatial regions for which spectra were extracted, as described in the text.}
\label{fig:profiles}
\end{centering}
\end{figure*}

\begin{figure*}
\begin{centering}
\includegraphics[width=8cm,angle=0]{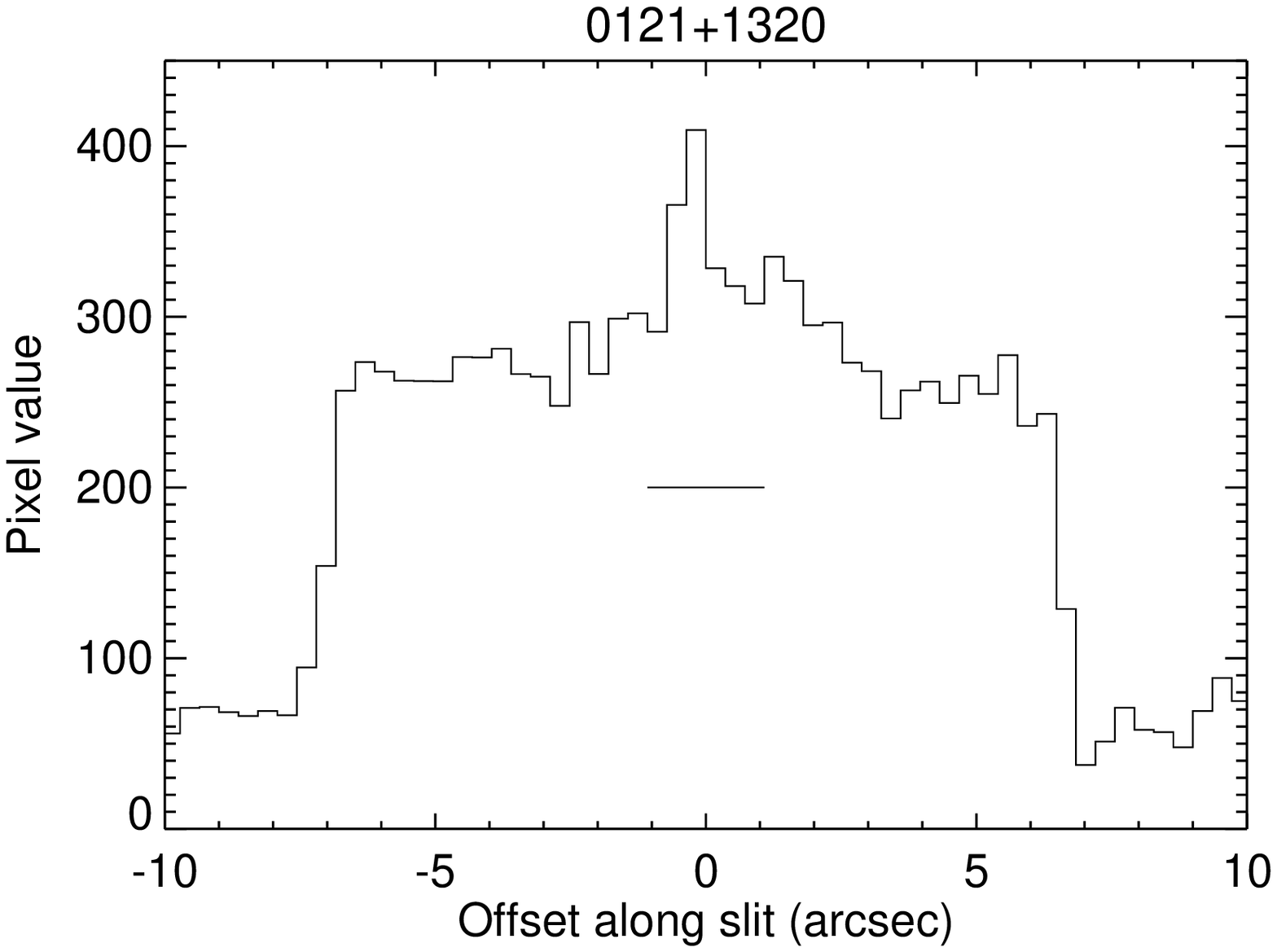}
\includegraphics[width=8cm,angle=0]{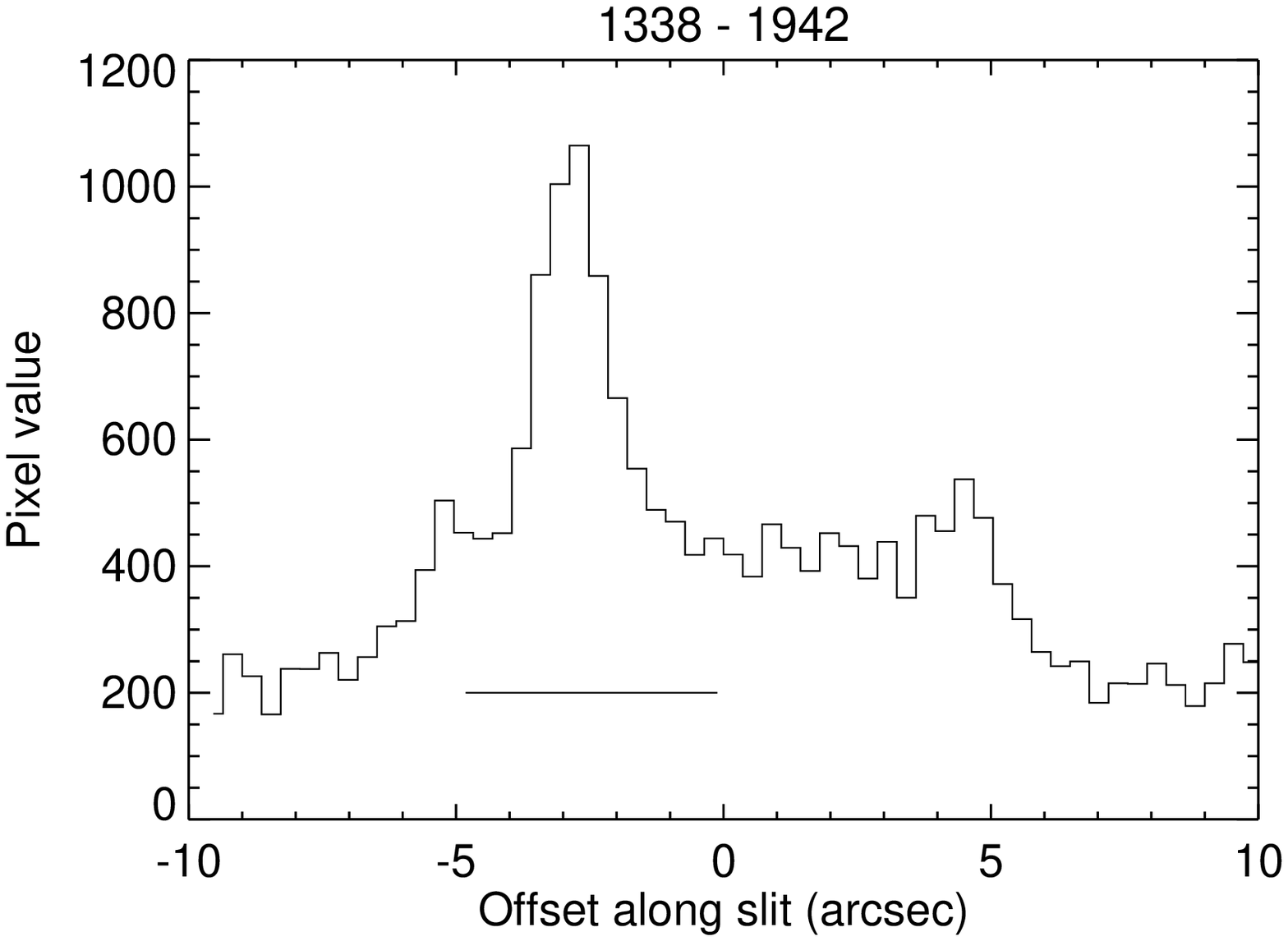}
\includegraphics[width=8cm,angle=0]{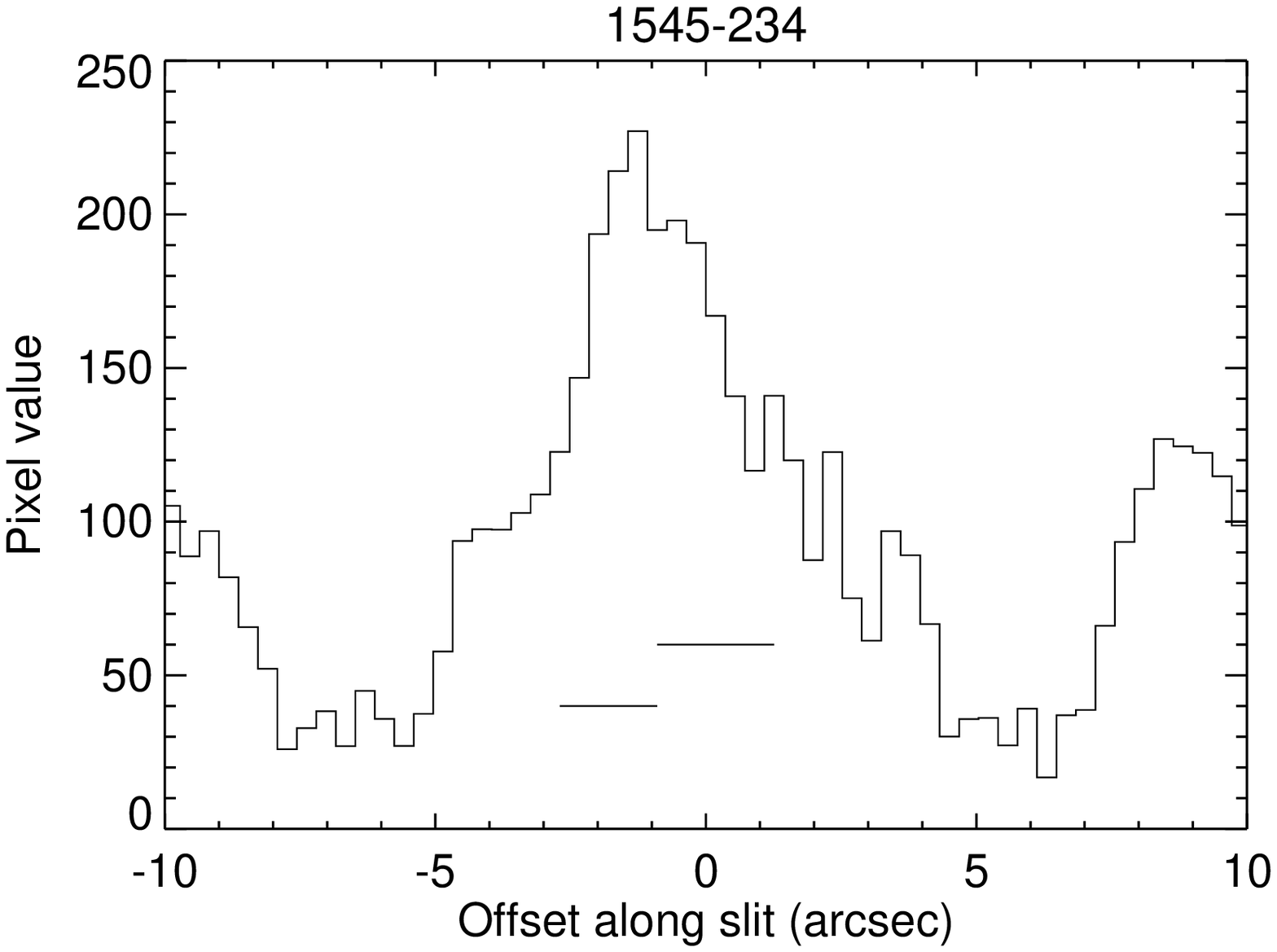}
\includegraphics[width=8cm,angle=0]{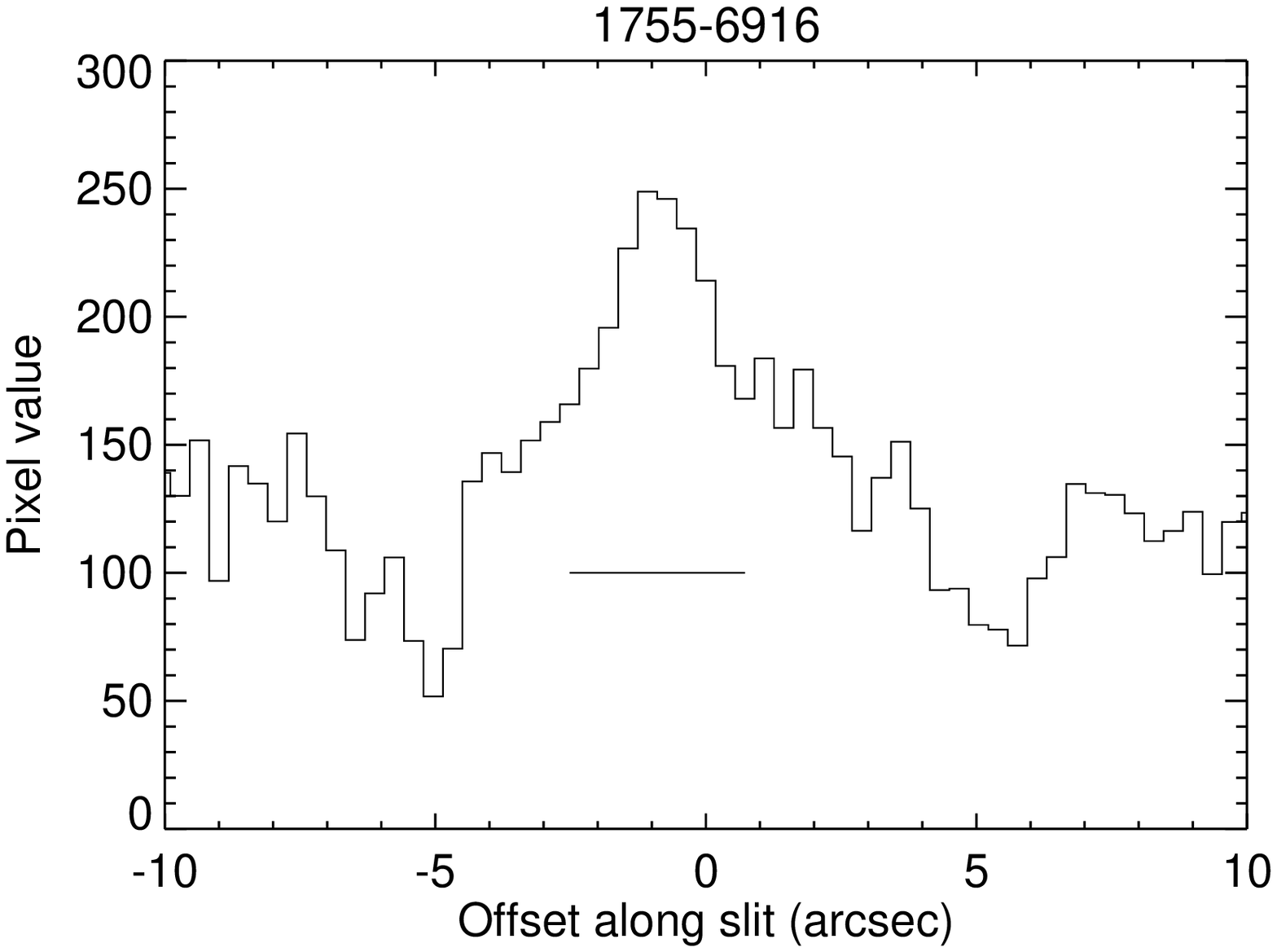}
\includegraphics[width=8cm,angle=0]{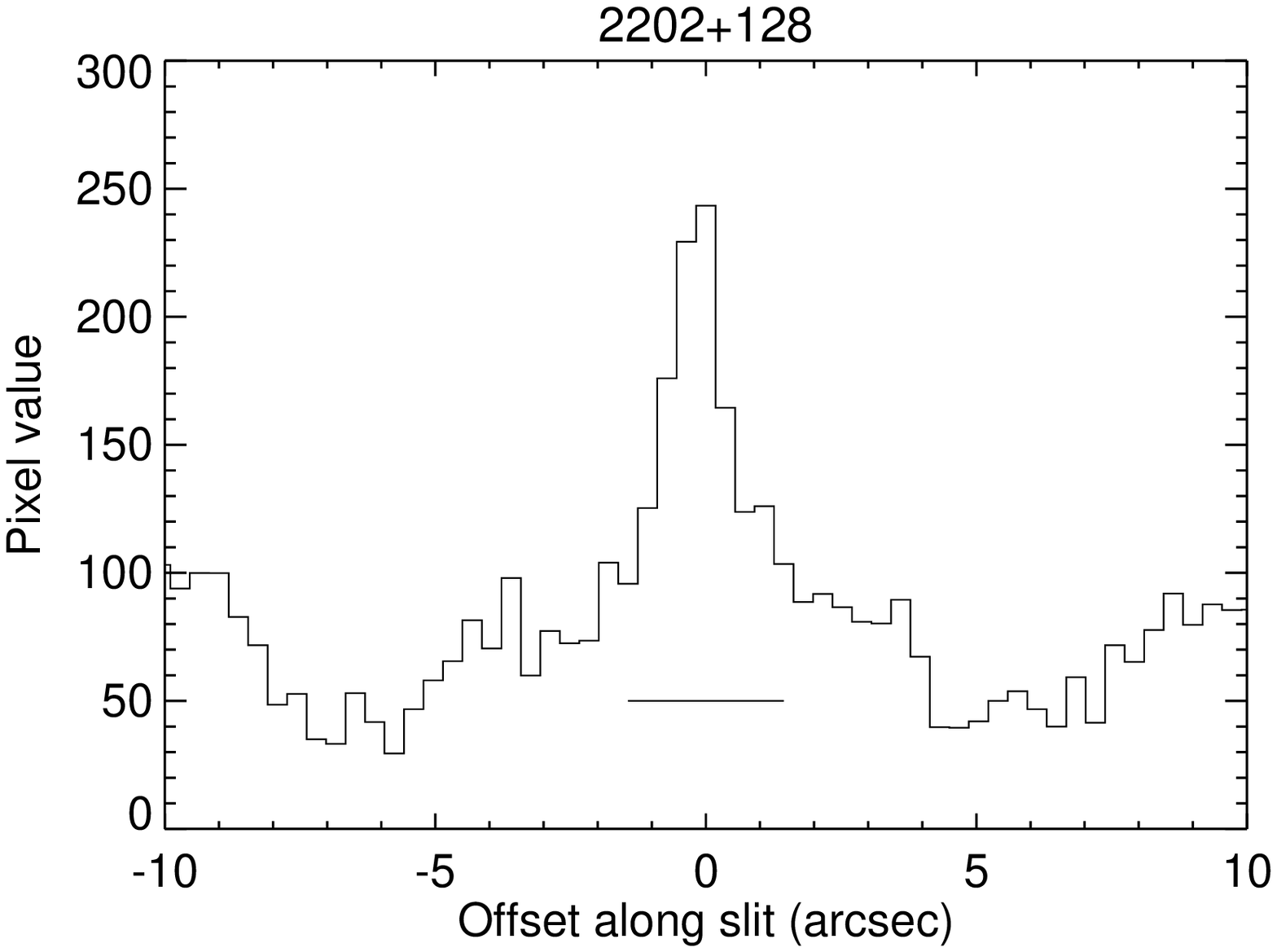}
\caption{Profiles of \Lya~along the 10~arcsec-long slit, formed by summing over 6 or 11 pixels in the dispersion direction (depending on the strength of the emission) near the peak in the observed emission. The gaps between the orders are clearly visible. The horizontal bars show the regions from which the spectra shown in Fig.~\ref{fig:profiles} were extracted; the regions used for background subtraction are not shown. }
\label{fig:spprofs}
\end{centering}
\end{figure*}

The radio galaxies and details of their profile fits are described separately below. The profiles are shown together in Fig.~\ref{fig:profiles} and the fitted parameters tabulated in Table~2. The quoted uncertainties on the fitted parameters denote the range of $\Delta \chi^{2}=1$, and were obtained by stepping the parameter 
of interest through its best-fit value and repeating the fit at each point. They should be regarded as minimum errors because systematic effects probably dominate over formal fitting errors. The major sources of the latter are: (i) the assumption that the underlying emission envelope is symmetric; (ii) the fact that the number of absorption profiles to be fitted is decided upon by visual inspection, and in cases of when lines overlap the Voigt profile decomposition is not unique; (iii) the degeneracy between $b$ parameter and $N_{\rm{HI}}$ on the flat part of the curve of growth where the line core is saturated. Fig.~\ref{fig:spprofs} shows spatial profiles of the \Lya~emission along the slit, to give an indication of its spatial extent.

To facilitate a more direct comparison with future simulations, we have extracted from the fits values for the mean transmitted flux $<F>$ beneath the emission envelope, i.e. $<F>$ is the ratio of the observed \Lya~flux to that in the underlying emission envelope which we assume would be observed in the absence of absorption. The values are shown in Table~3 and are given separately for the whole line and for the two halves of the line (defined relative to the centre of the emission envelope). In some cases, noise on the data can lead to $<F>$ values in excess of unity.

\begin{table*}
\caption{Absorption model fits to the UVES \Lya~spectra}
\begin{tabular}{lllll} \hline
Radio galaxy  & \Lya~emission redshift    & Absorption redshift   & Column density              & Doppler b parameter  \\
              & ($z_{\rm{em}}$)      &  ($z_{\rm{abs}}$)          &  log($N_{\rm{HI}}$) (\psqcm) & (\kmps) \\ \hline
0121+1320     & 3.522 $\pm$ 0.002      &  3.518 $\pm$ 0.0001 & 18.43 $\pm$ 0.02 $\star$ & 105 $\pm$ 6 $\star$ \\
              &             &  3.532 $\pm$ 0.0002 & 13.66 $\pm$ 0.09 & 84 $\pm$ 18 \\ \hline
1338-1942     & 4.105 $\pm$ 0.008           &  4.096 $\pm$ 0.001 & 15.02 $\pm$ 0.02 & 232 $\pm$ 4 \\
              &             &  4.087 $\pm$ 0.0001 & 19.67 $\pm$ 0.03 & 76 $\pm$ 4 \\ \hline
1545--234     & 2.751 $\pm$ 0.0009          &  2.753 $\pm$ 10$^{-5}$ & 13.89 $\pm$ 0.02 & 27 $\pm$ 1.5 \\
(component 1) &             &  2.751 $\pm$ 10$^{-5}$ & 13.74 $\pm$ 0.04 & 51 $\pm$ 3.5 \\
              &             &  2.749 $\pm$ 10$^{-5}$ & 14.65 $\pm$ 0.05 & 61 $\pm$ 2  \\
              &             &  2.746 $\pm$ $2 \times 10^{-5}$ & 14.39 $\pm$ 0.02 & 102 $\pm$ 3.5 \\
(component 2) & 2.752 $\pm$ 0.001           &  2.753 $2 \times 10^{-5}$ & 13.88 $\pm$ 0.02 & 34 $\pm$ 1 \\
              &             &  2.752 $\pm$ $2 \times 10^{-5}$ & 14.04 $\pm$ 0.04 & 38 $\pm$ 2 \\
              &             &  2.749 $\pm$ $2 \times 10^{-5}$ & 14.68 $\pm$ 0.03 & 76 $\pm$ 3.5 \\
              &             &  2.746 $\pm$ $4 \times 10^{-5}$ & 14.30 $\pm$ 0.02 & 92 $\pm$ 5.5 \\ \hline
1755--234     & 2.548 $\pm$ 0.001      &  2.546 $\pm$ $10^{-5}$  & 13.80 $\pm$ 0.004 & 41 $\pm$ 0.2 \\
              &             &  2.541 $\pm$ $10^{-5}$ & 14.09 $\pm$ $0.02$ & 45 $\pm$ 4 \\
              &             &  2.551 $\pm$ $2 \times 10^{-5}$ & 13.70 $\pm$ $0.05$ & 14 $\pm$ 0.5 \\ \hline
2202+128      & 2.705 $\pm$ 0.001           &  2.702 $\pm$ $2 \times 10^{-5}$ & 14.17 $\pm$ 0.01 & 58 $\pm$ 1 \\
              &             &  2.699 $\pm$ $2 \times 10^{-5}$ & 13.18 $\pm$ 0.04 & 20 $\pm$ 2 \\
              &             &  2.696 $\pm$ $2 \times 10^{-5}$ & 14.11 $\pm$ 0.02 & 30 $\pm$ 2 \\
              &             &  2.703 $\pm$ $6 \times 10^{-5}$   & 15.03 $\pm$ 0.01 & 67 $\pm$ 2 \\ \hline
0200+015$\dagger$      & 2.230  $\pm$ 0.04    &  2.2239 $\pm$ 0.0007 & 14.96 $\pm$ 0.06 & 612 $\pm$ 70 \\
              &             &  2.2282 $\pm$ 0.0002  & 14.73 $\pm$ 0.14  & 116 $\pm$ 27 \\
              &             &  2.2307 $\pm$ 0.0001 & 14.58 $\pm$ 0.16 & 61 $\pm$ 8 \\
              &             &  2.2349 $\pm$ 0.0001 & 14.38 $\pm$ 0.10 & 86 $\pm$ 11 \\ \hline
0943--242$\dagger$     & 2.922  $\pm$ 0.01     &  2.9066 $\pm$ 0.0062 & 14.02 $\pm$ 0.30 & 88 $\pm$ 45 \\
              &             &  2.9185 $\pm$ 0.0001 & 19.08 $\pm$ 0.06 & 58 $\pm$ 3 \\
              &             &  2.9261 $\pm$ 0.0005 & 13.55 $\pm$ 0.16 & 109 $\pm$ 35 \\
              &             &  2.9324 $\pm$ 0.0001 & 13.35 $\pm$ 0.13 & 23 $\pm$ 17 \\ \hline \hline
\end{tabular}
%ERRORS ON FITTED PARAMETERS YET TO BE CALCULATED
\begin{minipage}{170mm}
$\star$ Due to the absence of damping wings, the absorber could equally well be described as a superposition of one or
more lines from the flat part of the curve of growth, i.e. $log N_{\rm{HI}} \simeq 15- 18$, as discussed in the text. \\
$\dagger$ From our pilot study (see Jarvis et al.~2003). \\
\end{minipage}
\end{table*}

\begin{table}
\caption{Mean transmitted \Lya~flux for the HzRGs}
\begin{tabular}{|lllll|} \hline
Target  &  z & $<F>$       & $<F>$  & $<F>$  \\
        &    & (blue wing)$\dagger$ & (red wing)  & (whole line) \\ \hline
0121+1320 &  3.517 &  0.55 & 0.74 & 0.64  \\ 
1338-1942 &  4.106 &  0.17 & 1.0 & 0.60  \\
1545-234 (comp 1)&   2.755 & 0.40 & 0.79 & 0.59    \\
1545-234 (comp2) & 2.755 & 0.40 & 0.92 & 0.66 \\
1755-6916 &  2.55 & 0.80 & 1.10 & 0.93     \\
2202+128 &  2.704 & 0.51 & 1.0 & 0.76     \\ 
0943-242 &  2.93 &  0.61 & 0.75 & 0.68    \\
0200+015 &  2.23 &  0.29 & 0.66 & 0.48    \\ \hline
mean & -  & 0.47 & 0.87 & 0.67 \\ \hline
\end{tabular}
\begin{minipage}{76mm}
$\dagger$ Red and blue wings are defined relative to the centroid of the underlying emission envelope
\end{minipage}
\end{table}

\subsection{Notes on individual objects}

\subsection*{0121+1320}
This radio galaxy was discovered by De Breuck et al.~(2001) via the ultra-steep spectrum (USS) method for finding high-redshift radio galaxies. Even in the low resolution Keck discovery spectrum, a clear HI absorption system is visible in the \Lya. Furthermore, CO has been detected in emission in this object at the same redshift as the HI absorption (De Breuck et al.~2003). The same had been previously seen in the radio galaxy 2330+3927 at $z=3.087$, with the mass of H$_{\rm{2}}$ inferred from the CO exceeding that in HI and HII by a factor 100--1000 (De Breuck et al.~2002,2003). The UVES spectrum reveals 2 HI absorbers, the main absorber with a column density in excess of $10^{18}$\psqcm~and a much weaker system on the red wing of the line. We note, however, that the column density of the main system is still on the relatively flat part of the curve of growth (i.e. with a saturated core but before the damping wings have set in) so the line could equally well be described by a much lower $N_{\rm{HI}} \simeq 10^{16}$\psqcm~with correspondingly higher $b$ parameter ($b=170$ \kmps), or by a superposition of several lines with lower $b$. The CO detection does, however, favour the higher HI column quoted in Table~2.

\subsection*{1338--1942}
This radio galaxy was also discovered by De Breuck et al.~(2001) and found to be one of the most luminous \Lya~emitters 
in its class. It was subsequently shown to be part of a proto-cluster of 20 \Lya-emitting galaxies lying up to 
1.3~Mpc away (Venemans et al.~2002), and the same narrow-band image showed that the \Lya~halo of the radio galaxy itself is 
extremely asymmetric, extending by over 10\arcsec~to the northwest. The peak of the \Lya~and rest-frame optical emission are 
coincident with the brighter hotspot of an extremely asymmetric radio source. The low resolution ($FWHM \sim 5.5$\AA) VLT 
FORS1 spectrum of De Breuck et al.~(1999) shows that the \Lya~has an extremely asymmetric spectral profile, cutting off 
steeply on the blue side. They interpreted this as evidence for HI absorption with a column density in the range 3.5--$13 \times 
10^{19}$\psqcm, but residual flux in the absorption trough suggests additional complexity which higher resolution UVES 
spectroscopy might reveal.

The UVES \Lya~spectrum shown in Fig.~\ref{fig:profiles} was extracted from a region spanning 4.7\arcsec~along the slit. 
Although the \Lya~emission is clearly spatially extended on this scale, there is no obvious spatial structure to either the 
emission or the absorption. A 2-component absorption model is preferred over a single absorber, with the stronger absorber having 
a column density at the lower end of the range found by De Breuck et al.~(1999) for a single absorber. To search for velocity structure in the emission and absorption gas, we extracted a spectrum separately for each half of the 4.7\arcsec-wide region shown in Fig.~\ref{fig:profiles}, i.e. with a separation of 2.35\arcsec~between their centroids ($\equiv 16.4$\kpc~in the assumed cosmology). Fits with two absorbers were performed on each spectrum and the velocity difference (in the frame of the radio galaxy) between the peaks of the underlying envelopes found to be $103 \pm 55$\kmps~(the velocity differences between the two regions for the strong and weak absorbers are $54 \pm 128$ and $70 \pm 33$\kmps, respectively). Under the naive assumption that the velocity gradient in the emission peaks arises from symmetric Keplerian motion, the mass enclosed within the central 8.2\kpc~would be $5 \pm 2.7 \times 10^{9}$\Msun. The stellar mass in this region is, however, much greater, $\sim 10^{12}$\Msun (inferred from the K-band magnitude of De Breuck et al.~1999 using the galaxy evolution tracks in Rocca-Volmerange et al.~2004), indicating that the gas dynamics are in fact dominated by forces other than gravity, e.g. the radio jet.

\subsection*{1545--234}
The \Lya~spectrum obtained by van Ojik et al. (resolution $\sim 2.8$\AA) exhibits a single absorption system on the blue wing of the line, with $\log N_{\rm{HI}}=18.2 \pm 0.3$ and $b=39 \pm 10$\kmps. Their 2-dimensional spectrum shows that the absorption does not cover the entire extent of the \Lya~emission. In contrast, the UVES spectrum reveals considerable additional complexity as we identify instead 4 distinct absorption systems at substantially lower column density. Since the emission has a spatial extent of 4\arcsec~($\equiv 32$\kpc), we extracted spectra for 2 distinct spatial regions, each covering 2\arcsec~along the slit (components 1 and 2 in Table 2), but there is no evidence for any significant spatial variation in the properties of the absorbers. The absorption redshifts are identical within the errors and the column densities vary by less than a factor of 2. This constitutes further evidence that the absorbing shells are coherent structures with lateral extents of at least several tens of \kpc.

\subsection*{1755--6916}
This radio galaxy was also discovered by De Breuck et al.~(2001) who report that the \Lya~emission is spatially extended by over 7\arcsec~with a complex morphology. The UVES spectrum shows that the \Lya~profile is distinctly non-gaussian in appearance, with a central peak and prominent wings: a lorenzian profile was thus used to characterise the underlying emission, upon which we fit three absorption systems with $\log N_{\rm{HI}} \simeq 14$. The non-gaussian profile of the underlying emission could be due to resonant scattering of the \Lya~emission, or it might instead be a genuinely broad base from a relatively unobscured quasar broad line region. Spectroscopy of the H$\alpha$ emission line is needed to distinguish between these alternatives.

\subsection*{2202+128}
The \Lya~spectrum in van Ojik et al. (resolution $\sim 1.7$\AA) exhibits a strong absorption system just blueward of the peak 
($\log N_{\rm{HI}} \simeq 18$), together with two weaker absorbers on the red wing of the line ($\log N_{\rm{HI}} \simeq 14.7$), one
of which takes the form of a `shoulder' with a large Doppler parameter ($b \sim 350$\kmps). The much higher resolution of UVES 
once again uncovers additional complexity: the main absorber is better fit by a model with 2 absorbers of column densities $\log N_{\rm{HI}} \sim 14$ and 15. There is no strong evidence for the other two absorbers identified by van Ojik et al. but we do now see two weak absorbers blueward of the main absorber, one of which is hinted in the van Ojik et al. spectrum but not modelled due to the poor signal to noise. Since the absorbers are thought to arise on scales of many kpc, we clearly do not expect to see any varibility on 10 year timescales. Slit placement differences offer a more likely explanation for any significant difference in absorption structure in the two spectra.

\subsection*{Pilot study targets: 0200+015 and 0943--242}
In Table 2 and the statistical analysis of the next section, we also include the two targets from our UVES pilot study (Jarvis et al.~2003; Wilman et al.~2003), the main results of which are summarised in section 1.

\section{STATISTICAL PROPERTIES OF THE ABSORPTION LINE ENSEMBLE}
Having fitted an absorption model to the \Lya~line of each radio galaxy, we now examine the statistical properties of the 
ensemble of absorbers. Firstly, we simply compare the number of HI absorbers seen within the FWHM of the underlying 
emission profile for each line, with the number expected from the column density distribution function of the \Lya~forest 
seen in quasar spectra. For the latter we use the parameterisation given by Kim et al.~(1997) which, at $z \sim 2.85$, takes 
the form $f(N_{\rm{HI}}) = 4.9 \times 10^{7}N_{\rm{HI}}^{-1.46}$; $f(N_{\rm{HI}})$ gives the number of HI systems per unit 
column density per unit redshift path, $X(z)$ (for the $q_{\rm{0}}=0$ cosmology they use, $X(z) \equiv 0.5 
[(1+z)^{2} - 1]$). For our current purposes, this power-law is an adequate approximation to the distribution function for 
the entire column density range of interest, $N_{\rm{HI}}=13-20$, although there is some evidence that it steepens at higher 
redshift (exponent $-1.55$ at $z \sim 3.7$) and deviates below it for $\log N_{\rm{HI}} > 15$ at low-redshift 
($z \sim 2.3$). We consider the column density intervals $\log N_{\rm{HI}}=13-15$ and $\log N_{\rm{HI}} > 18$, but for the 
former we exclude from the calculation regions occupied by $\log N_{\rm{HI}} > 18$ absorbers, against which 
$\log N_{\rm{HI}}=13-15$ absorbers could clearly not be detected. The results are are shown in Table~4. 

There we see that the incidence of strong absorption ($\log N_{\rm{HI}} > 18$) is much higher than expected from the 
statistical properties of the intergalactic medium (IGM) at large. This much we knew from the work of van Ojik et al. In the 
lower column density regime, the issue is more ambiguous: in the high redshift objects 0121+1320 ($z=3.517$) and 1338-1942 
($z=4.106$) we see fewer lines than in the IGM at these redshifts, whilst the opposite is true for the targets at 
$z=2-3$. We emphasise that this is simply an empirical comparison between the number of systems observed in each line and 
the number expected over such a wavelength interval from the IGM at these redshifts. The applicability of this distribution 
function in the present case, given that radio galaxies reside in large overdensities and induce peculiar motions in their 
vicinity (quite apart from the effects of photoionisation and outflows from the central object), is discussed in section 5.

\begin{figure}
\includegraphics[width=0.35\textwidth,angle=270]{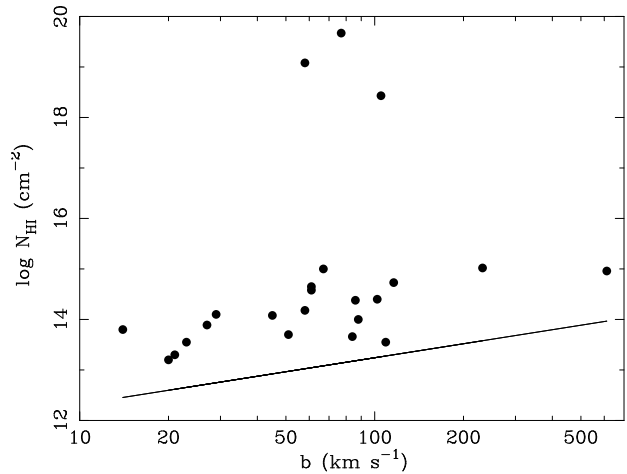}
\caption{\normalsize The relation between the Doppler b parameter and column density for the HI absorbers. The line shows the 
form of the optical depth selection function, in this case for $\tau=0.1$.}
\label{fig:b_nh}
\end{figure}

\begin{figure}
\includegraphics[width=0.35\textwidth,angle=270]{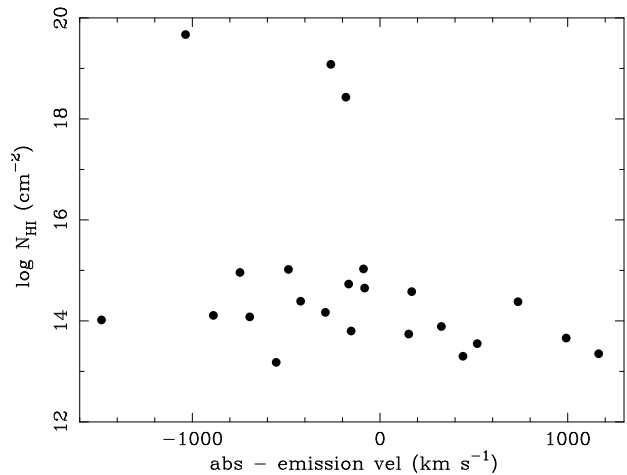}
\caption{\normalsize The HI column density of the absorbers plotted against its velocity with respect to peak of the underlying \Lya~emission, evaluated in the rest frame of the radio galaxy.}
\label{fig:voff}
\end{figure}

In Fig.~\ref{fig:b_nh}, we show the relation between the Doppler b parameter and the HI column density for the absorbers.
Those in the interval $\log N_{\rm{HI}}=13-15$ occupy the same range of b parameter as the \Lya~forest clouds in Kim et al., 
i.e. 10--100\kmps. In part, this is a selection effect -- we cannot resolve widths below $\sim 10$\kmps, and the optical
depth selection ensures that the minimum detectable column density increases linearly with the b parameter.  
Fig.~\ref{fig:voff} shows that these absorbers are spread over the entire width of the emission line profile: the mean 
velocity offset from the emission peak is $-80$\kmps, but the dispersion in velocity offset is much larger at 640\kmps. In 
contrast, all three of the high column density ($\log N_{\rm{HI}} > 18$) systems are blueshifted from the emission peak; such 
a tendency had been seen in the larger, lower resolution sample of van Ojik et al. Figs.\ref{fig:b_nh} and \ref{fig:voff} 
also illustrate an important fact not immediately obvious from Table~2, namely that there is a gap of 3 orders of magnitude 
in column density, $\log N_{\rm{HI}} \simeq 15-18$ in which we do not detect any absorption systems. This is not, however, 
inconsistent with the statistics for the IGM, as such systems are a factor of $\sim 10$ less numerous than $\log N_{\rm{HI}}=13-15$ systems, so with reference to Table~2 there would be roughly a 10 per cent chance of seeing one against any given 
\Lya~emission line. We do, however, re-iterate the caveat that a line with $\log N_{\rm{HI}} \simeq 18$ can, however, be 
described as a superposition of absorbers from the flat part of the curve of growth ($\log N_{\rm{HI}} \simeq 15-18$) with a 
range of $b$ parameters (although 0121+1320 is the only one of our objects to which this is applicable).

\begin{table}
\caption{Comparison with the statistics of the \Lya~forest for the IGM}
\begin{tabular}{|llllll|} \hline
Target  &  z & $13 \leq log(N_{\rm{HI}}) \leq 15$  & $\log(N_{\rm{HI}}) > 18 $ \\
        &  & Observed (expected) $\dagger$ &	Observed (expected) \\ \hline
0121+1320 &  3.517 & 1 (1.8)             &  1 (0.005)       \\ 
1338-1942 &  4.106 & 1 (3.3)             &  1 (0.006)       \\
1545-234 &   2.755 & 4 (1.0)	         &  0 (0.006)       \\
1755-6916 &  2.55 & 3 (1.6)            &  0 (0.009)       \\
2202+128 &  2.704 & 4 (1.6)              & 0 (0.009)        \\ 
0943-242 &  2.93 & 3 (1.2)              & 1 (0.01)        \\
0200+015 &  2.23 & 4 (1.6)              & 0 (0.009)        \\
\hline
\end{tabular}
\begin{minipage}{76mm}
Column densities in units of \psqcm.\\
$\dagger$ Compares observed number of HI absorption systems in \Lya~line with number expected
from the column density distribution function of QSO absorbers. See text.
\end{minipage}
\end{table}

\section{DISCUSSION}

We now consider the implications of our results for models of the origin and evolution of the absorption systems and their 
connection to the radio source and the \Lya~forest. In Jarvis et al.~(2003) we interpreted the results from our pilot study in 
the context of the model put forward by Binette et al., in which the absorbers comprise shells of low density gas, subject 
to disruption by the expanding radio source and metal enrichment by starburst superwinds. Not withstanding these sources of 
influence, the clear assumption was that these high column density shells ($\log N_{\rm{HI}} > 18$) ultimately had no causal connection with the radio source itself and formed independently of it, as a by-product of massive galaxy formation. If this were the case, 
we would expect to see such absorbing shells around young, similiarly massive galaxies at these epochs, prior to the ignition of 
any nuclear radio source. The most promising candidates for such a population are the SCUBA galaxies. They exhibit the same K--z relation as the radio galaxies (Serjeant et al.~2003) and indeed there is a sizeable overlap between the two populations (Archibald et al.~2001; Reuland, R\"{o}ttgering \& van Breugel~2003). The spectroscopic evidence concerning the existence or otherwise of absorbing haloes around (non-radio-galaxy) SCUBA sources is, however, inconclusive. Chapman et al.~(2003) have recently obtained Keck spectra of 34 SCUBA sources and find that 10 of them exhibit emission line spectra with a median redshift of $z=2.4$, but the spectral resolution ($\sim 8$\AA) and signal-to-noise level are insufficient to infer the presence of anything but the strongest absorption (although a few per cent of SCUBA source do show hints of absorbed \Lya; Smail, private communication). Higher resolution spectra of these sources are required. Also worthy of such scrutiny are the radio-quiet X-ray type II quasars, particularly those known to have narrow emission-line spectra similar to those of radio galaxies. Examples of the latter are the $z=3.7$ source in the Chandra Deep Field South (Norman et al.~2002) and the $z=3.3$ source in the Lynx Field (Stern et al.~2002), but from existing low resolution spectroscopy it is not clear whether their narrow \Lya~emission lines have associated absorption. Larger samples of type II quasars are becoming available for high resolution spectroscopy from X-ray surveys (e.g. Mainieri et al.~2002; Gandhi et al.~2004) and the the Sloan Digital Sky Survey (Zakamska et al.~2004).

Likewise, SAURON integral field spectroscopy (Bower et al.~2004) shows no evidence for any strong \Lya~absorption against Steidel `blob 1', an extended \Lya~halo in the $z=3.11$ SSA22 supercluster, or either of its two nearby Lyman break galaxies (absorbers with $\log N_{\rm{HI}} =13-15$ cannot be detected in these spectra due to resolution and signal:noise limitations). The Steidel blobs bear a striking resemblance to the \Lya~haloes of HzRGs, but lack central radio sources, although `blob 1' does have an associated SCUBA source. The fact that they were discovered by virtue of their high equivalent width \Lya~emission is a strong selection effect against absorption. The Lyman break galaxies exhibit a broad distribution of \Lya~equivalent width centred at $\sim 0$\AA~(Shapley et al.~2003), ranging from objects with strong \Lya~emission to galaxies with strongly damped absorption. An example of the latter is the $z=2.724$ gravitationally-lensed galaxy MS1512--cB58 which has absorption with $N_{\rm{HI}} \simeq 6 \times 10^{20}$\psqcm~(Savaglio, Panagia \& Padovani~2002). The velocity offsets between the emission and absorption lines (stellar and interstellar) in Lyman break galaxies (e.g. in MS1512--cB58, Pettini et al.~2000, 2002; see also Shapley et al.~2003) suggest that the absorbing gas is in the form of galactic outflow driven by the high star formation rate. An HII galaxy somewhat less massive than the LBGs is the $z=3.357$ gravitationally-lensed Lynx arc (Fosbury et al.~2003), which exhibits an absorbing wind with comparable column densities in HI and CIV ($\simeq 10^{15}$\psqcm), attributed to outflows from $10^{6}$ O stars.

We conclude that more observations are needed before we can make a definitive statement about the existence or otherwise of 
absorbing shells around high-redshift galaxies (not quasars) which are comparable in mass to the HzRG, but which lack 
radio-loud AGN. For this reason, it remains possible that the absorbing shells we see in the HzRGs are a generic by-product of
massive galaxy formation, as proposed by Binette et al. and assumed by Jarvis et al. There are, however, other models in which the origin of the shells is directly connected to the radio source, which we now discuss.

\subsection{The expanding radio source and the origin of high column density absorption}
Jarvis et al.~dismissed the possibility that the absorbing gas might take the form 
of `shells propagated by the radio source', on the grounds that such shells would be: (i) disrupted at radii less than the 
extent of the radio source; (ii) incapable of producing the observed line-of-sight absorption, given that the radio sources 
are oriented roughly along the plane of the sky. For such a scenario these reasons are clearly valid. 

There is, however, a more natural way in which a radio source can give rise to shells of absorbing gas: a highly supersonic 
radio jet expanding into the dense medium of a young galaxy will be surrounded by an advancing quasi-spherical bow-shock. 
Immediately behind this shock, swept-up gas is heated to a few times $10^{7}$\K~and radiatively cools to a few times $10^{4}$\K, 
at which point an appreciable fraction becomes neutral and could be responsible for the observed absorption. This idea was 
proposed and investigated in detail by Krause~(2002). In his radiative hydrodynamical simulations, after a few million 
years of radio source propagation the shell will provide a neutral column of $3.8 \times 10^{21}$\psqcm~around the whole radio 
source, with 98 per cent of the matter displaced by the bow-shock being contained within the shell (the remaining 2 per cent, 
or $\simeq 10^{9}$\Msun, is entrained within the cocoon and could contribute to the observed line emission; the hotter material 
immediately behind the bow-shock contributes negligibly to the observed levels of \Lya~emission in HzRGs). The velocity width 
of the absorbing gas is approximately its internal sound speed, 55\kmps~on average in his simulation. The shells are, however, 
subject to a number of instabilities. Firstly, the shells are thermally unstable and should fragment into individual sub-shells, 
which would be preferentially blue-shifted from the main absorber, but almost never by more than 250\kmps. Secondly, the shells 
are also gravitationally unstable, a process which will, he claims, result in the formation of roughly $10^{4}$ globular systems 
each of $10^{6}-10^{7}$\Msun~(in line with the globular cluster excess seen locally in central cluster galaxies compared with 
other elliptical galaxies of the same luminosity). As the shells fragment, their covering factor against the background emission 
will drop and eventually fall to zero, accounting for the absence of absorption among the larger radio sources in the van Ojik et 
al. sample.

\begin{figure}
\includegraphics[width=0.48\textwidth,angle=0]{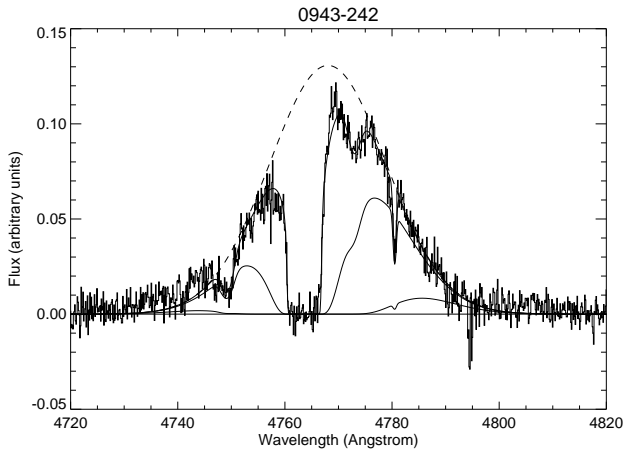}
\caption{\normalsize The \Lya~profile of 0943--242 with 3 absorption models overlaid (from top to bottom): the best 
fit model from Jarvis et al. where the main absorber has $\log N_{\rm{HI}}=19.1$; with $N_{\rm{HI}}$ for this 
absorber alone increased by factors of 10 and 100, respectively. The dashed line is the emission envelope.}
\label{fig:0943sim}
\end{figure}

To what extent can Krause's model explain our observations? Firstly, the spatially-extended nature of the absorbers is consistent with the quasi-spherical bow-shock of the model. It is clear, however, that HI column densities comparable to his 
predicted level of $3.8 \times 10^{21}$\psqcm~could not be seen against \Lya~emission in these sources: the damping wings would
remove essentially all the emission and render the line undetectable, as Fig.~\ref{fig:0943sim} illustrates for the case of 0943--242. Therefore, our measurements of column densities $\log N_{\rm{HI}} \simeq 18-20$ reflect a selection bias towards systems 
with lower column densities, due either to environmental effects or evolution, perhaps due to the initial 
stages of shell fragmentation. We note, however, that 21~cm radio observations, which offer a continuum against which to search
for HI absorption, can in principle be used to probe the higher $N_{\rm{HI}}$ regime. Such a search by Vermeulen et 
al.~(2003) in a sample of compact (i.e. young) radio sources out to $z=0.85$ revealed absorbing columns of 
$\log N_{\rm{HI}}=19-21.5$ (for a covering factor of unity and spin temperature 100\K). The distribution of 21~cm absorption 
velocity (with respect to optical redshift) is skewed to negative values, with some sources offset by more than 1000\kmps, as we 
find for our strongest absorber (in 1338--242). These authors argue, however, that the absorbing material resides in a compact disc-like structure much closer to the nucleus that the \Lya~absorbers in the HzRGs. Furthermore, attempts to detect 21cm absorption in HzRGs have been largely unsuccessful (R\"{o}ttgering et al.~1999), with 0902+34 at $z=3.4$ being the only source with a solid detection (Uson et al.~1991; Briggs et al.~1993; de Bruyn et al.~1995).

As remarked by Jarvis et al., any model for the absorption systems in HzRGs must also be applicable to radio-loud quasars, 
by virtue of the orientation-based unification schemes for the two populations. We therefore compare with the work
of Baker et al.~(2002) on associated CIV absorption in radio-loud quasars. The incidence of such absorption is higher than 
in radio-quiet quasars and is preferentially found in steep spectrum and lobe-dominated quasars, implying that the absorbing 
material lies away from the radio-jet axis (where it would be strongly influenced by the line of sight UV continuum). The 
absorption strength anti-correlates with the projected linear size of steep spectrum radio sources. All this is consistent 
with our results on HzRG absorption and with the Krause model. But using a sample of more luminous radio-loud and radio-quiet 
quasars, Vestergaard~(2003) cast doubt on some of the conclusions of Baker et al., and ascribed the differences to luminosity 
effects. In general, however, the lack of information on the spatial extent of the absorbers in quasars leaves open the 
possibility that some of their absorption arises on much smaller scales than in the Krause model, and is more directly 
associated with the central engine through mechanisms similar to those cited for broad absorption line (BAL) quasars 
(e.g. disk-wind models). For very high-redshift ($z \geq 5$) quasars, Barkana \& Loeb~(2003) have suggested that HI absorption 
can be produced as gas cools behind the accretion shock associated with its collapse onto the host dark matter halo. However, 
since their fits to the \Lya~profiles are poor and the optical depth for such absorption scales as $(1+z)^4$, we do not consider
this to be important for our HzRGs.

\subsection{The lower column density systems: fragments of larger shells?}
But what of the absorbers with $\log N_{\rm{HI}} \simeq 13-15$, are they formed by fragmentation of the larger shells? The 
simulations of Krause do not have the resolution to follow the evolution of the shells into this regime. His arguments 
for fragmentation stem from the application of analytical instability criteria. The shells are thermally unstable since 
the sound crossing time through the newly-formed shell ($\simeq 5 \times 10^{5}$\yr) exceeds the cooling time 
($\simeq 10^{4}$\yr). Thus on the latter timescale, one would expect the shells to fragment by a factor of $\sim 50$ in 
column density [gravitational instability proceeds much more slowly, on timescales of a few $10^{7}$\yr]. However, it 
is not clear how the {\em neutral} column density will evolve as the shells fragment. For a uniform slab in 
photoionization equilibrium with a diffuse UV radiation field, the neutral fraction will decrease rapidly as the degree 
of fragmentation increases, and eventually drop to zero when the shell thickness becomes comparable to the Str\"{o}mgren 
depth. In the present case, however, we are dealing with a collisionally ionized, cooling (i.e. non-equilibrium) plasma 
to which these arguments are not applicable. Therefore, short of performing detailed simulations, we cannot exclude the 
possibility that some of the $\log N_{\rm{HI}} \simeq 13-15$ absorbers are fragments of initially larger shells. 

However, for the following reasons we believe that low column density absorbers are a distinct population, not formed by 
fragmentation of the larger shells: (i) there appear to be no absorbers in the 3 order of magnitude wide gap 
from $\log N_{\rm{HI}} \simeq 15-18$, an absence which would be surprising unless fragmentation were very rapid in this 
regime; (ii) the velocities of the $\log N_{\rm{HI}} \simeq 13-15$ absorbers bear no obvious relation to the higher 
$N_{\rm{HI}}$ systems -- Krause expects the fragments to be preferentially blue-shifted, but by no more than 250\kmps; 
(iii) with respect to the \Lya~emission redshift, the $\log N_{\rm{HI}} \simeq 13-15$ absorbers show a velocity dispersion 
comparable to that expected for a proto-cluster~(Fig.~\ref{fig:voff}); (iv) the $\log N_{\rm{HI}} \simeq 13-15$ 
absorbers follow the same $b-N_{\rm{HI}}$ relation as the IGM \Lya-forest (Fig.~\ref{fig:b_nh}); (v) the observed number 
of $\log N_{\rm{HI}} \simeq 13-15$ systems in each galaxy is within a factor of 2--4 (at most) of the number expected in 
the IGM at large (Table~4). We therefore believe that the low column density absorbers are part of the population that 
gives rise to the \Lya~forest in the IGM, albeit modified by the massive galaxy/protocluster environment, as discussed 
below.

\subsection{The \Lya~forest in the proto-cluster environment}
There is now direct evidence that high-redshift radio galaxies reside in proto-cluster environments: narrow-band 
\Lya~imaging and follow-up spectroscopy has confirmed the existence of overdensities of \Lya~emitters within 
$\sim 3$\Mpc~of 6 HzRGs at $z=2-4$, by factors 4--15 with respect to blank fields (Kurk et al.~2000; Pentericci et 
al.~2000; Venemans et al.~2002).  The velocity dispersions of the proto-clusters decrease from $\sim 1000$\kmps~at $z=2$ 
to $\sim 300$\kmps~at $z=4$ (Kurk et al.~2003a). The overdensity around 1338-242 at $z=4.1$ is very irregular, with the radio 
galaxy situated off-centre, suggesting that the structure has barely separated from the Hubble flow. In contrast, the $z=2.16$ 
proto-cluster around 1138-262 shows evidence for segregation between the various galaxy populations (\Lya-emitters, H$\alpha$ 
emitters and EROs), indicative of an older, more dynamically relaxed structure (Kurk et al.~2003b).

%\begin{table}
%\caption{Mean transmitted \Lya~flux for the HzRGs}
%\begin{tabular}{|lllll|} \hline
%Target  &  z & $<F>$       & $<F>$  & $<F>$  \\
%        &    & (blue wing)$\dagger$ & (red wing)  & (whole line) \\ \hline
%0121+1320 &  3.517 &  0.55 & 0.74 & 0.64  \\ 
%1338-1942 &  4.106 &  0.17 & 1.0 & 0.60  \\
%1545-234 (comp 1)&   2.755 & 0.40 & 0.79 & 0.59    \\
%1545-234 (comp2) & 2.755 & 0.40 & 0.92 & 0.66 \\
%1755-6916 &  2.55 & 0.80 & 1.10 & 0.93     \\
%2202+128 &  2.704 & 0.51 & 1.0 & 0.76     \\ 
%0943-242 &  2.93 &  0.61 & 0.75 & 0.68    \\
%0200+015 &  2.23 &  0.29 & 0.66 & 0.48    \\ \hline
%mean & -  & 0.47 & 0.87 & 0.67 \\ \hline
%\end{tabular}
%\begin{minipage}{76mm}
%$\dagger$ Red and blue wings are defined relative to the centroid of the underlying emission envelope
%\end{minipage}
%\end{table}

As discussed in section 1, cosmological hydrodynamical simulations predict an increase in the opacity of the \Lya~forest 
in the vicinity of massive galaxies (e.g. Croft et al.~2002; Bruscoli et al.~2003, for z=3). The 
predictions are usually expressed in terms of the (continuum-normalised) mean transmitted flux, $<F>=e^{-\rm{\tau}}$, 
where $\tau$ is \Lya~optical depth. At large transverse distance from a galaxy ($>6$\Mpc), $<F>$ is predicted to be 
close to the IGM average of 0.67, but to fall with decreasing impact parameter to values below 0.2 within 0.5\Mpc, 
even when the effects of galaxy winds are taken into account (Bruscoli et al.~2003). Allowance for observational 
uncertainty in redshift determination can increase the minimum $<F>$ by a factor $\sim 2$, but not enough to match the
results of Adelberger et al.~(2003), which show that $<F>$ increases to 0.9 within 0.5\Mpc~of $z=3$ Lyman-break galaxies. 
We note that Savaglio et al.~(2002) analysed the Ly$\alpha$ forest towards the $z=2.724$ Lyman break galaxy MS1512-cB58, and
found evidence for excess absorption with respect to QSO sightlines at a co-moving distance of $\sim 140$\Mpc. However, the 
presence of strongly damped \Lya~absorption prevented them from analysing the properties of the forest closer into the galaxy.

Our results cannot be interpreted in the same theoretical framework as those of Adelberger et al. because our absorption 
statistics are in redshift space along the line-of-sight, not in real space perpendicular to it. At $z=2,3$ and 4 the Hubble 
parameter has values of $\simeq 210$, 310 and 430\kmpspMpc, respectively, so given an intrinsic \Lya~linewidth of $\sim 1500$\kmps~we are sensitive to (at the most) $\sim 7,5$ and 3.5\Mpc~in redshift space, and around half these values if the \Lya~peak is at the systemic velocity since we can only see absorption from foreground material. On such scales around a massive galaxy ($M_{\rm{total}}=1.5 \times 10^{11}$\Msun/$h$) at $z=3$, peculiar velocities will be up to 150\kmps~(Bruscoli et al.~2003). It will be of interest to compare our average $<F>$ values given in Table~3 (as well as \Lya~profiles and probability density 
functions of transmitted flux) with new simulations of the \Lya~forest in the vicinity of massive galaxies {\em along the 
line-of-sight}, and at a variety of epochs. The intriguing possibility remains that the simulations may reveal that all the 
\Lya~absorption in HzRGs can be accounted for by the \Lya~forest, with the strong absorbers ($N_{\rm{HI}}>10^{18}$\psqcm) in 
fact being a superposition of several weaker systems. 

The observed evolution in proto-cluster properties (velocity dispersion, dynamical segregation) from $z=4$ to $z=2$, suggests 
that for our lower-z targets it may be more appropriate to think of the HI absorption as a probe of the proto-intracluster 
medium, rather than of the IGM (although at some point the distinction is purely semantic). Zhan, Fang \& Burnstein~(2003) have 
recently argued that the proto-intracluster medium at $z>2$ is likely to be significantly multiphased and similar to the 
spiral-rich group environment at lower redshift. At such epochs supernova heating is insufficient to heat the ICM to 
$10^{6}-10^{8}$\K~globally, and it more likely consists of pockets of metal-rich gas at $10^{4}-10^{5}$\K. From the lowest value 
of the transmitted flux found by Adelberger et al.~(2003), $<F>=0.52$ at $\simeq 2$\Mpc~from Lyman-break galaxies and assuming 
a path length for this absorption of $12h^{-1}$\Mpc, Zhan et al. infer an upper limit on the HI column density of 
$N_{\rm{HI}}=6.4 \times 10^{14}$\psqcm~through the ICM at $z=3$. Our measured column densities are therefore consistent with a 
multiphase ICM at these redshifts.

\section{CONCLUSIONS}
We have built on our successful pilot study and increased the number of high-redshift radio galaxies with UVES echelle 
spectroscopy of \Lya~from 2 to 7. With this enlarged ensemble of absorption systems we can begin to make some inferences 
about their origin. The new spectra demonstrate once again that HI absorption is more complex when seen at this much higher 
resolution: several absorbers identified in low-resolution spectra with $N_{\rm{HI}}>10^{18}$\psqcm~now fragment into a 
number of much weaker systems with $N_{\rm{HI}}<10^{15}$\psqcm. There appears to be a gap in the $N_{\rm{HI}}$ distribution, as 
we identify strong absorbers with $N_{\rm{HI}} \simeq 10^{18}-10^{20}$\psqcm~and weaker systems with $N_{\rm{HI}} \simeq 
10^{13}-10^{15}$\psqcm, but none at intermediate $N_{\rm{HI}}$. 

We discussed the origin of the strong absorption systems with $N_{\rm{HI}}>10^{18}$\psqcm~in the framework put forward 
by Krause~(2002), based on radio source physics. A highly supersonic jet expanding into a dense proto-galactic 
medium will be surrounded by a quasi-spherical bow shock. The simulations by Krause~(2002) show that the cooling of post-shock 
material will lead to neutral column densities comparable to those observed, and that due to instabilities the shells 
will fragment and offer much less absorption towards larger radio sources. The model provides a very natural explanation for the 
observation that the shells have high covering factors and surround the entire galaxy. Appealing though it is, it would be
premature to enshrine it as the canonical model for the strong absorbers in HzRGs. Several new challenging observations are needed
before we can determine whether such strong absorbers really are produced by the radio source bow shock or whether they are 
instead a by-product of massive galaxy formation. Firstly, in the HzRGs themselves deep, high resolution spectra should be 
obtained of the faint kinematically-quiescent \Lya~haloes, identified by Villar-Martin et al.~(2003). Secondly, such spectroscopy 
should also be obtained of a carefully selected sample of comparably massive high-redshift galaxies without radio-loud AGN, 
comprising SCUBA galaxies and narrow-line X-ray type II quasars.

We cannot exclude the possibility that some of the lower column density systems with $N_{\rm{HI}}=10^{13}-10^{15}$\psqcm~form by 
fragmentation of larger shells until the simulations of the bow-shock absorbers are extended into this regime. The rate of 
incidence of such absorption in our spectra and their distribution on the $b$--$N_{\rm{HI}}$ plane suggests, however, that these 
systems could be part of the IGM \Lya~forest population. Simulations predict that the opacity of the forest should increase 
within a few Mpc~of a massive galaxy at these epochs. Our results show an enhancement (by factors 2--4) in the number of 
\Lya~forest systems (relative to the IGM) at $z=2-3$ and a deficit by similar factors in the two $z>3.5$ systems, although on 
the grounds of small number statistics we do not claim that this implies evolution over this redshift interval. For a proper 
comparison, independent of any Voigt fitting uncertainties and systematics, new simulations of the \Lya~forest should be 
performed. Unlike those in the literature which predict the HI opacity at the galaxy redshift as a function of transverse 
distance from it, the new simulations should yield the \Lya~forest along the line of sight to a massive object in redshift space, 
as a function of galaxy mass and epoch. In lieu of this, we note that the observed column densities are consistent with simple 
estimates for a multiphase proto-intracluster medium at $z>2$.

\section*{ACKNOWLEDGMENTS}
RJW and MJJ acknowledge financial support from PPARC. The new data presented in this paper were obtained at the European Southern Observatory, Chile (Programme ID:71.B-0036(A)).

{}

\end{document}